\documentclass[aps,
twocolumn,
superscriptaddress]{revtex4}

\usepackage{amsfonts}
\usepackage{amsmath}
\usepackage{amssymb}
\usepackage{graphicx}%
\usepackage{hyperref}
\usepackage[dvipsnames]{xcolor}
\usepackage{ulem}

\begin{document}

\title{Electron correlation effects in paramagnetic cobalt}

\author{A. S. Belozerov}
\affiliation{Department of Physics, University of Hamburg, 20355 Hamburg, Germany}

\author{V. I. Anisimov}
\affiliation{M. N. Mikheev Institute of Metal Physics, Russian Academy of Sciences, 620108 Yekaterinburg, Russia}
\affiliation{Ural Federal University, 620002 Yekaterinburg, Russia}




\begin{abstract}
We study the influence of Coulomb correlations on spectral and magnetic properties of fcc cobalt using a combination of density functional theory and dynamical mean-field theory. The computed uniform and local magnetic susceptibilities obey the Curie-Weiss law, which, as we demonstrate, occurs due to the partial formation of local magnetic moments. We find that the lifetime of these moments in cobalt is significantly less than in bcc iron, suggesting a more itinerant magnetism in cobalt. In contrast to the bcc iron, the obtained electronic self-energies exhibit a quasiparticle shape with the quasiparticle mass enhancement factor ${m^*/m}\sim$1.8, corresponding to moderately correlated metal. Finally, our calculations reveal that the static magnetic susceptibility of cobalt is dominated by ferromagnetic correlations, as evidenced by its momentum dependence.
\end{abstract}


\maketitle

{\bf Introduction.} Metallic cobalt is a canonical ferromagnet with an extremely high Curie temperature of 1418~K, which is significantly larger than that in iron (1043~K) and nickel (631~K). 
The experimental magnetic moment of ferromagnetic cobalt is also high (1.7~$\mu_{\rm B}$) and exceeded only by iron (2.2~$\mu_{\rm B}$) among all $3d$ metals.
These characteristics make cobalt an essential ingredient in a wide range of modern technological applications.

At low temperatures, cobalt is ferromagnetic with the hexagonal close packed (hcp) lattice, which upon heating to 720~K transforms into the face-centred cubic (fcc) one. 
Further heating to 1418 K leads to a transition to the paramagnetic phase, which is stable up to a melting point of 1770~K.

The electronic structure of cobalt has been theoretically studied using density functional theory (DFT)
within local density approximation (LDA) and generalized gradient approximation (GGA).
These studies addressed electronic, structural and magnetic properties of both phases~\cite{Moroni1997,Leung1991,Korling1992,Cho1996,Matar2007,shea2010}.
However, the LDA and GGA alone are known to have difficulties in the description of transition metals due to strong electron correlations in partially filled electronic subshells.
In a magnetically ordered state, these correlation effects can be partially treated by the static mean-field approximation within DFT+$U$ method~\cite{Anisimov1991}.
Indeed, the application of this method to ferromagnetic cobalt resulted in a better agreement with experimental data compared to GGA calculations~\cite{shea2010}.

However, the DFT+$U$ approach is not suitable for the paramagnetic state, and, in addition, it neglects the dynamic electron correlations, which were shown to be significant in other $3d$~metals~\cite{Katanin2010,OurGamma,ourChromium,Sangiovanni,ourVanadium}.
To obtain an accurate treatment of local many-body effects at finite temperatures, the DFT can be combined with other model approaches, such as the dynamical mean-field theory (DMFT)~\cite{Metzner1989,RMP1996}.
This theory neglects non-local correlation effects, assuming momentum-independent self-energy, and is exact in the limit of infinite coordination number.
The above-mentioned combination is called DFT+DMFT~\cite{dftdmft,Kotliar2006} and can be applied to both magnetically ordered and paramagnetic states at any ratio of Hubbard parameter~$U$ to bandwidth.

In this paper, we study the Coulomb correlation effects in paramagnetic fcc cobalt by the DFT+DMFT approach.
We demonstrate that fcc cobalt is a moderately correlated metal with partially formed local magnetic moments, the lifetime of which is 
substantially lower than in bcc iron,
being a system with well-formed local moments.

{\bf Method.}
We perform our study by a fully charge self-consistent DFT+DMFT approach~\cite{charge_sc2,charge_sc3} implemented with plane-wave pseudopotentials~\cite{espresso12,Leonov1}.
The exchange-correlation functional was considered within the Perdew-Burke-Ernzerhof form of GGA.
For an fcc lattice of cobalt, we 
adopt the equilibrium lattice constant of 6.731~a.u. obtained in our DFT+DMFT calculations.
This value is in good agreement with the experimental lattice constant of 6.714~a.u.~\cite{Armentrout2015}.
The convergence threshold for total energy was set to $10^{-6}$~Ry.
The kinetic energy cutoff for wavefunctions was set to 65~Ry.
The reciprocal space integration was performed using ${20\times 20\times 20}$\, ${\bf k}$-point grid except the calculations of momentum-dependent susceptibility, where ${60\times 60\times 60}$ grid was used.

Our DFT+DMFT calculations explicitly include the $3d$ and $4s$ valence states by constructing a basis set of atomic-centred Wannier functions~\cite{Wannier1,Wannier2,Wannier3} within the energy window spanned by the $s$-$d$ band complex.

For a parametrization of the on-site Coulomb interaction, we use Slater integrals $F^0$, $F^2$, and $F^4$ linked to the Hubbard parameter ${U\equiv F^0}$ and Hund's rule coupling ${J_{\rm H}\equiv (F^2+F^4)/14}$
(see Ref.~\cite{u_and_j}).
We perform our calculations with ${U=4}$~eV and ${J_{\rm H}=0.9}$~eV,
which are close to the estimates obtained by the constrained random-phase approximation~\cite{Aryasetiawan2006,Miyake1,Miyake2,Sasioglu} and constrained DFT calculations~\cite{Aryasetiawan2006}.
Moreover, these values
were widely used in DFT+DMFT studies of elemental iron~\cite{OurAlphaBelozerovKatanin,OurGammaKataninBelozerov,OurAlphaGammaTransition}, which is the neighbour of cobalt in the periodic table.

To take into account the electronic interactions already described by DFT, we use a double-counting correction in the around mean-field form, which is evaluated from the self-consistently determined occupations.
We also verified that the fully localized form of double-counting correction leads to similar results with a slightly smaller (by about 0.03) filling of \textit{d}~states.

The quantum impurity problem within DMFT was solved by hybridization expansion continuous-time quantum Monte Carlo method~\cite{CT-QMC_Rubtsov,CT-QMC_Werner} with the density-density form of Coulomb interaction.
The analytical continuation of self-energies to the real-energy range was performed by using Pad\'e approximants~\cite{Pade}.

We note that the non-local correlation effects, neglected within DMFT, are expected to be weak for the considered fcc lattice, due to its relatively large coordination number.

{\bf Electronic properties.}
Our DFT+DMFT calculations yield the $d$-states filling of 7.67, which is about one electron larger than that in iron.
This fact favours weaker many-body effects in cobalt compared to iron, since the strongest electron correlations are expected near the half-filling of electronic subshells.

\begin{figure}[t]
\centering
\includegraphics[clip=true,width=0.44\textwidth]{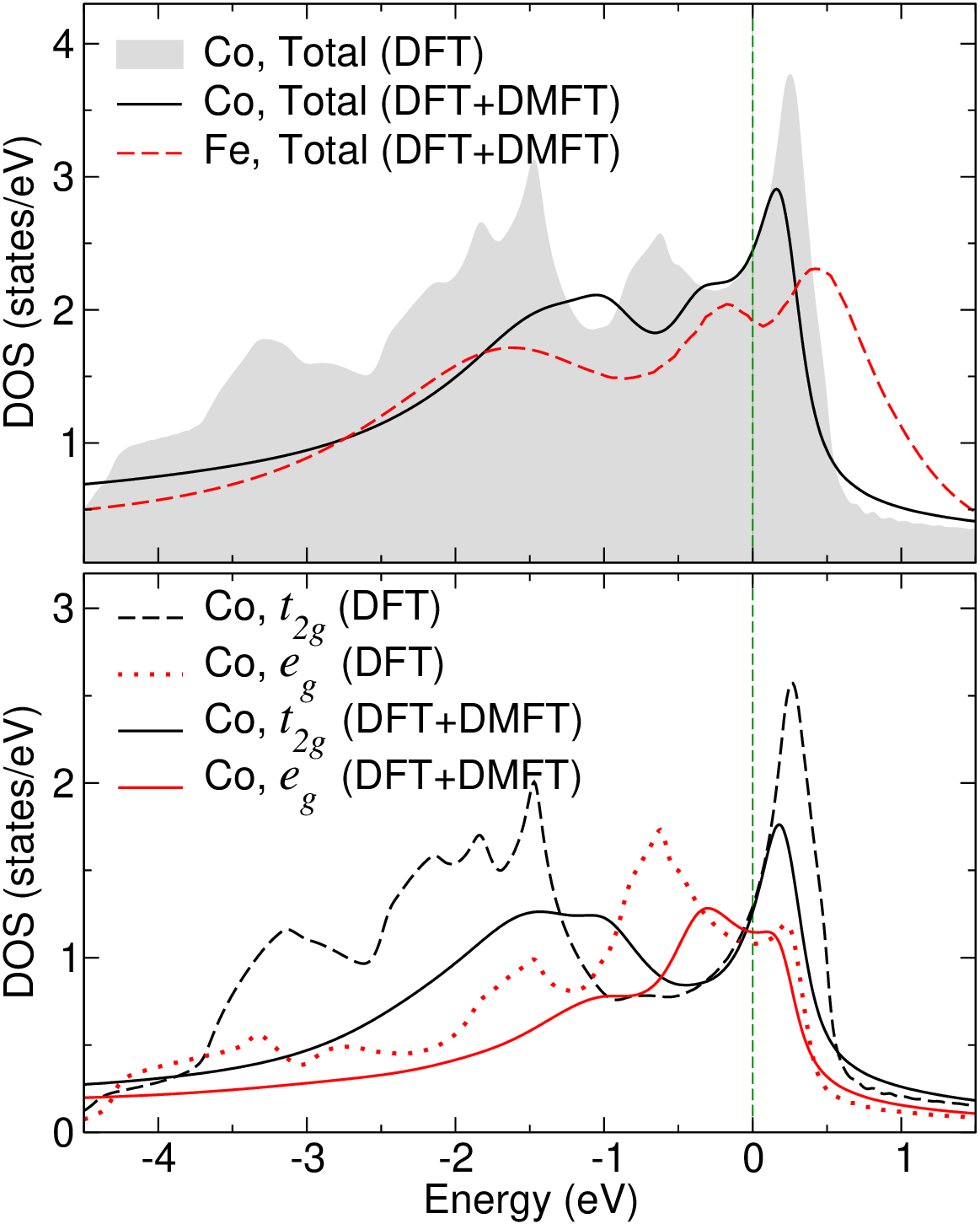}
\caption{
\label{fig:dos}
Fig. 1. Total (top panel) and orbital projected (bottom panel) density of states obtained by non-magnetic DFT and DFT+DMFT for cobalt (both panels) and bcc iron~\cite{OurAlphaBelozerovKatanin} (top panel) at temperature ${T = 1658}$~K. The Fermi level is at zero energy.}
\end{figure}

\begin{figure}[t]
\includegraphics[clip=true,width=0.45\textwidth]{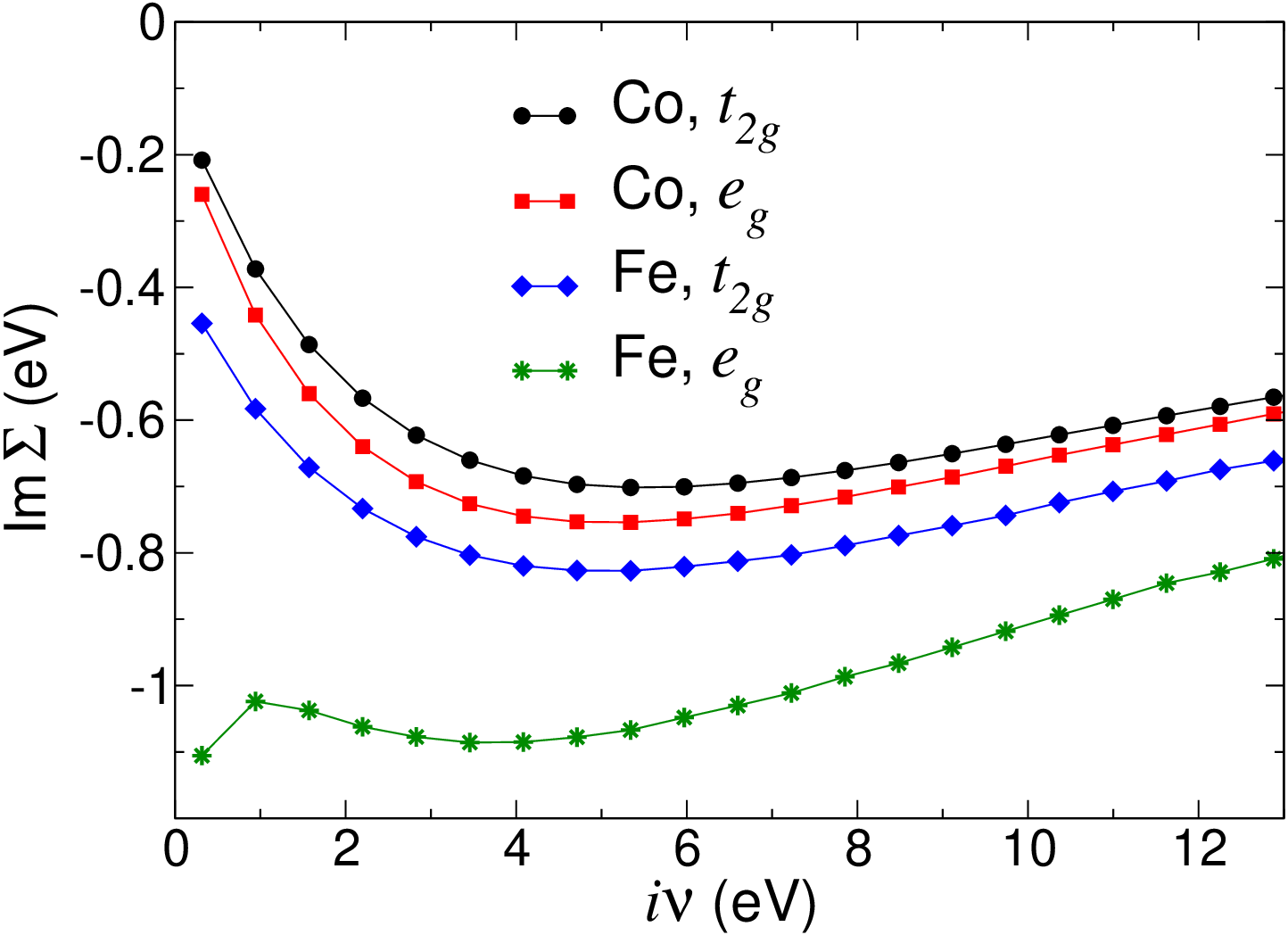}
\caption{
\label{sigma}
Fig. 2. Imaginary part of electronic self-energy as a function of Matsubara frequency $i\nu$ 
for cobalt and bcc iron~\cite{OurAlphaBelozerovKatanin} 
obtained by DFT+DMFT method at temperature of 1160~K.}
\end{figure}

In the top panel of Fig.~\ref{fig:dos}, we present the obtained total density of states (DOS) in comparison with that of bcc iron.
One can see that taking into account correlation effects within DMFT leads to the renormalization of Co DOS near the Fermi level, decreasing the distance from the Fermi level to the peak located above it.
A similar peak is observed in bcc iron, but it is further away from the Fermi level.
Most importantly, this peak in bcc iron, originating from the $e_g$ states, is probably responsible for the orbital-selective formation of local magnetic moments~\cite{Katanin2010}.
As shown in the bottom panel of Fig.~\ref{fig:dos}, the
peak in Co~DOS above the Fermi level originates from the $t_{2g}$ states.
Additionally, there is another peak located below the Fermi level. This peak is less prominent and comes from the $e_{g}$ states.

In Fig.~\ref{sigma} we display the imaginary parts of computed self-energy
$\Sigma(i\nu_n)$ as a function of Matsubara frequency~$\nu_n$.
The self-energies for cobalt have a quasiparticle shape, implying that they depend
on small $\nu_n$ as ${\textrm{Im}\, \Sigma(i\nu_n)\approx -\Gamma -(Z^{-1} {-}1)\nu_n}$, where $\Gamma$ is the quasiparticle damping and $Z$ is the quasiparticle residue.
This is in contrast to bcc Fe, where the states with $e_g$ symmetry show non-Fermi-liquid (non-quasiparticle) behaviour accompanied by formation of well-defined local magnetic moments~\cite{Katanin2010}.

To estimate the strength of electron correlation, we compute $Z^{-1}$,
which corresponds to the quasiparticle mass enhancement factor ${m^*/m}$ due to the locality of self-energy in DMFT. 
The resulting ${m^*/m}$ values in cobalt are 1.77 and 1.88 for $t_{2g}$ and $e_g$ states, respectively.
These values correspond to a moderately correlated metal
and are greater than those for chromium (1.17)~\cite{ourChromium}, nickel (1.25)~\cite{Sangiovanni}, and vanadium (1.7)~\cite{ourVanadium}.

\begin{figure}[t]
\centering
\includegraphics[clip=true,width=0.43\textwidth]{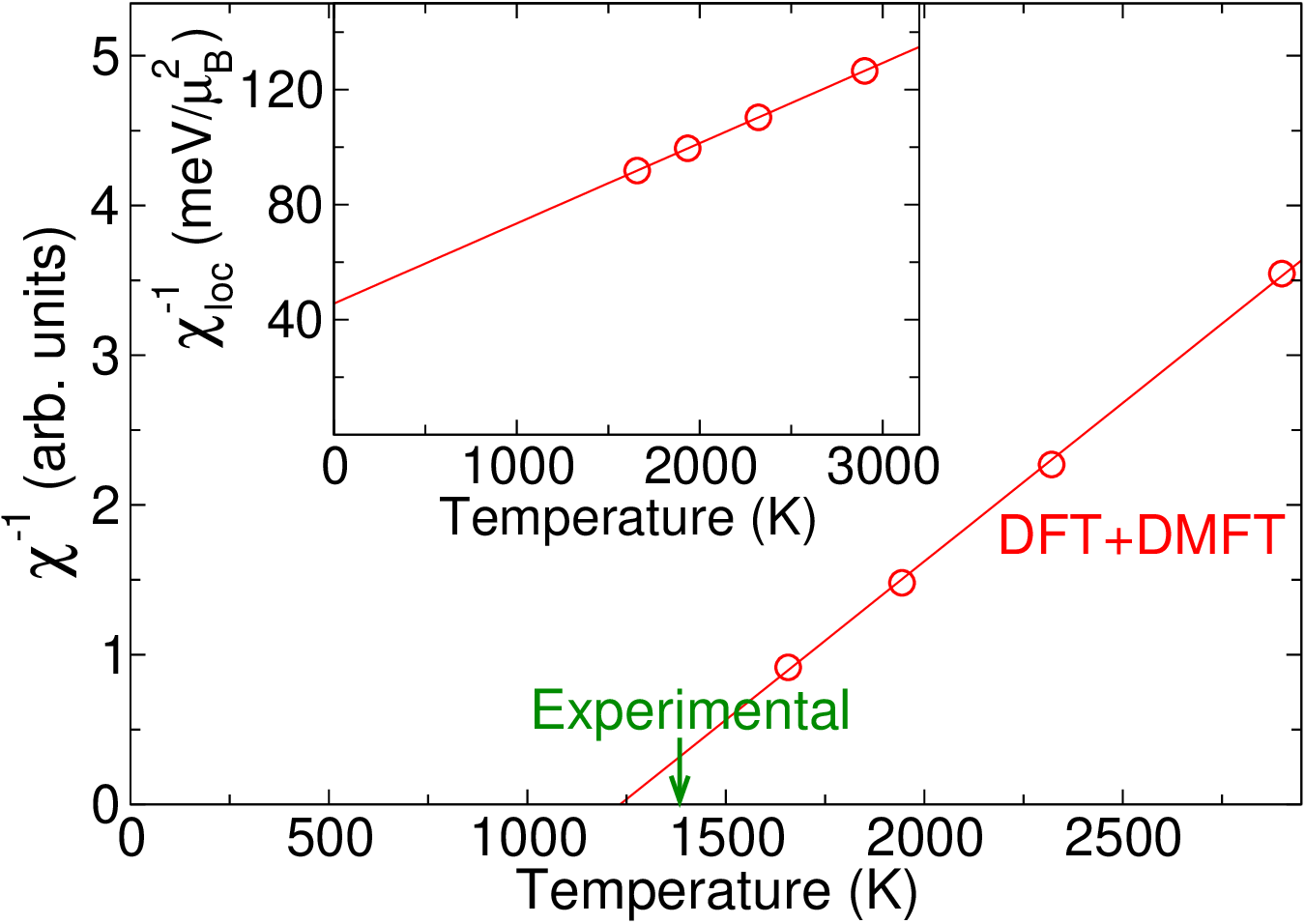}
\caption{
\label{Fig:chi_uniform}
Fig. 3. Inverse of uniform magnetic susceptibility $\chi$ (main panel) and inverse of local magnetic susceptibility $\chi_{\rm loc}$ (inset) as a function of temperature obtained by DFT+DMFT method for cobalt. The experimental value of Curie temperature is denoted by an arrow. The straight lines depict the least-squares fit to the linear dependence.}
\end{figure}

{\bf Magnetic properties.}
First, we compute the uniform magnetic susceptibility by applying a 
small external magnetic field, leading to the splitting of single-electron energies by 20~meV. We verified that the chosen magnetic field is small enough to provide a linear response. As shown in Fig.~\ref{Fig:chi_uniform}, the inverse of calculated uniform magnetic susceptibility follows the Curie-Weiss law at high temperatures. 
Performing linear extrapolation of inverse susceptibility, we extract the Curie-Weiss temperature ${\Theta=1240}$~K, which is 13\% less than the experimental Curie temperature of 1418~K.

To clarify the underlying cause of the Curie-Weiss behaviour, we calculate the local magnetic susceptibility $\chi_{\rm loc}=4\mu_{\rm B}^2 \int_0^\beta \langle S_z(\tau) S_z(0) \rangle d\tau$,
where $S_z$ is the $z$-component of the
local spin operator,
$\beta$ is the inverse temperature,
$\tau$ is the imaginary time.
The inverse of $\chi_{\rm loc}$, shown in the inset of Fig.~\ref{Fig:chi_uniform}, also depends linearly on temperature. This finding indicates that the Curie-Weiss behaviour of uniform magnetic susceptibility is caused by formation of local magnetic moments.

\begin{figure}[t]
\centering
\includegraphics[clip=true,width=0.485\textwidth]{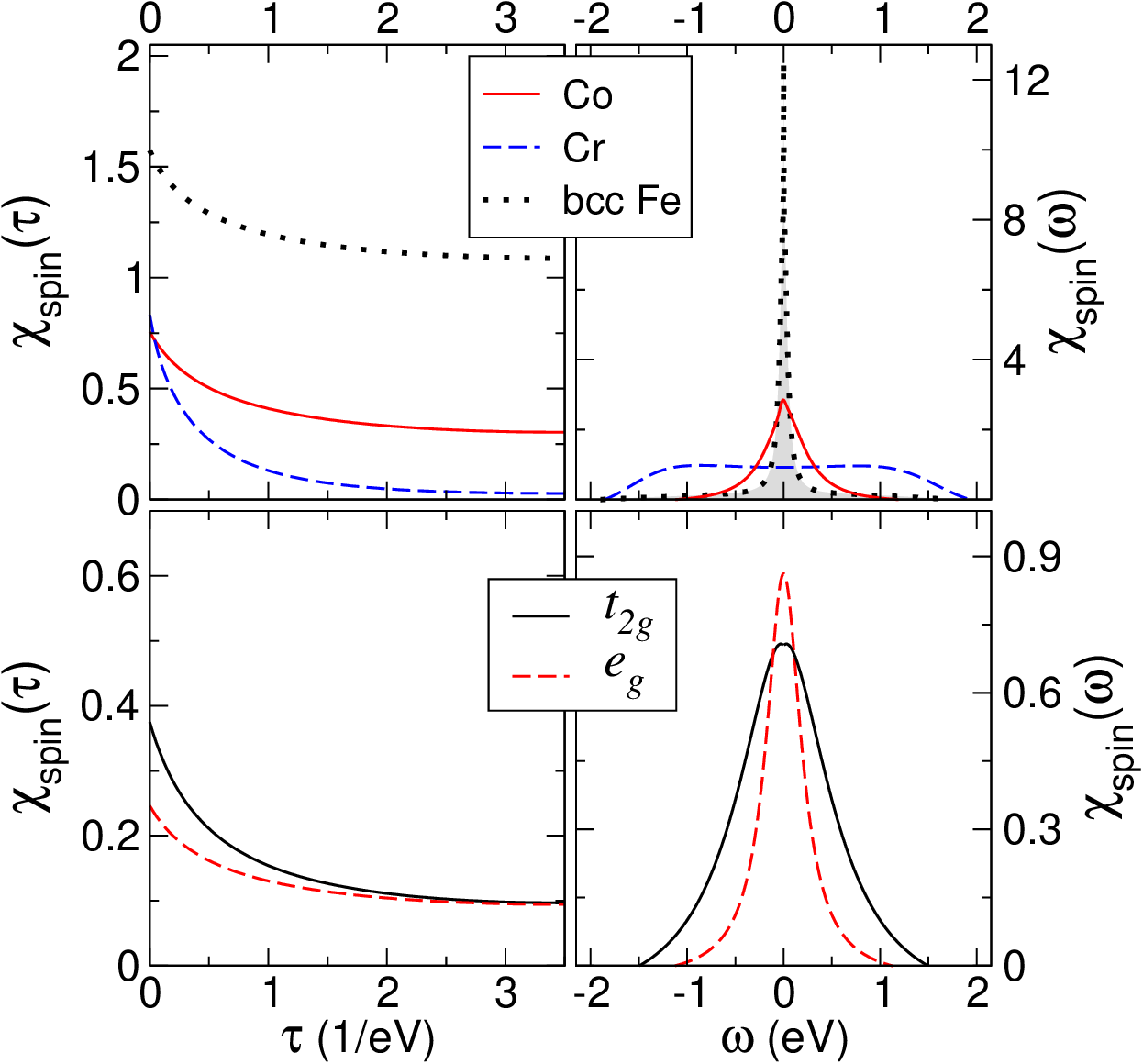}
\caption{
\label{Fig:correlators}
Fig. 4. Local spin-spin correlation functions in the imaginary time $\tau$ (left panels) and real frequency $\omega$ (right panels) domains calculated by DFT+DMFT method for cobalt 
in comparison with those of paramagnetic chromium~\cite{ourChromium} and bcc iron~\cite{OurAlphaBelozerovKatanin} (top panels).
Bottom panels: orbital-resolved spin-spin correlation functions for cobalt. 
All calculations are performed at temperature ${T=1658}$~K.
}
\end{figure}

To investigate the degree of magnetic moment localization, we examine the local spin-spin correlation function defined as ${\chi_{\rm spin}(\tau) = \langle S_z(\tau) S_z(0) \rangle }$ and its real-frequency counterpart $\chi_{\rm spin}(\omega)$.
The latter is obtained by Fourier transforming ${\chi_{\rm spin}(\tau)}$ to imaginary frequency and then analytically continuing to real frequency $\omega$ using Pad\'e approximants~\cite{Pade}.
In the top panel of Fig.~\ref{Fig:correlators}, we present the obtained ${\chi_{\rm spin}(\tau)}$ and the real part of ${\chi_{\rm spin}(\omega)}$ in comparison with those of paramagnetic chromium~\cite{ourChromium}, which is a canonical itinerant antiferromagnet, and bcc iron~\cite{OurAlphaBelozerovKatanin} known as a system with well-defined local magnetic moments~\cite{footnote1}.
One can see that ${\chi_{\rm spin}(\tau)}$ for cobalt has
an instantaneous average ${\langle S_z^2 \rangle = 0.75}$, which is
lower than iron's value of 1.6
and close to that of chromium.
However, in contrast to Cr, 
${\chi_{\rm spin}(\tau)}$ for cobalt saturates to a finite value at ${\tau\to\beta/2}$, indicating some localization of magnetic moments.

To obtain a more quantitative estimate of moments localization, we consider the real part of ${\chi_{\rm spin}(\omega)}$ shown in the right panels of Fig.~\ref{Fig:correlators}.
Specifically, the half-width of the peak in ${{\rm Re}[ \chi_\textrm{spin}(\omega)]}$ at its half-height is nearly equal to the inverse lifetime of local magnetic moments~\cite{OurGamma,Toschi3}.
The obtained results indicate that the lifetime of moments in cobalt is about 10 times lower than in bcc iron, but about 8 times larger than in chromium.

In order to clarify the orbital contributions to partially formed local moments, we present the orbital-resolved spin-spin correlation in the bottom panels of Fig.~\ref{Fig:correlators}.
The obtained results show that the lifetime of local magnetic moments in $e_g$ states is about twice larger than in $t_{2g}$ ones.

\begin{figure}[t]
\centering
\includegraphics[clip=true,width=0.44\textwidth]{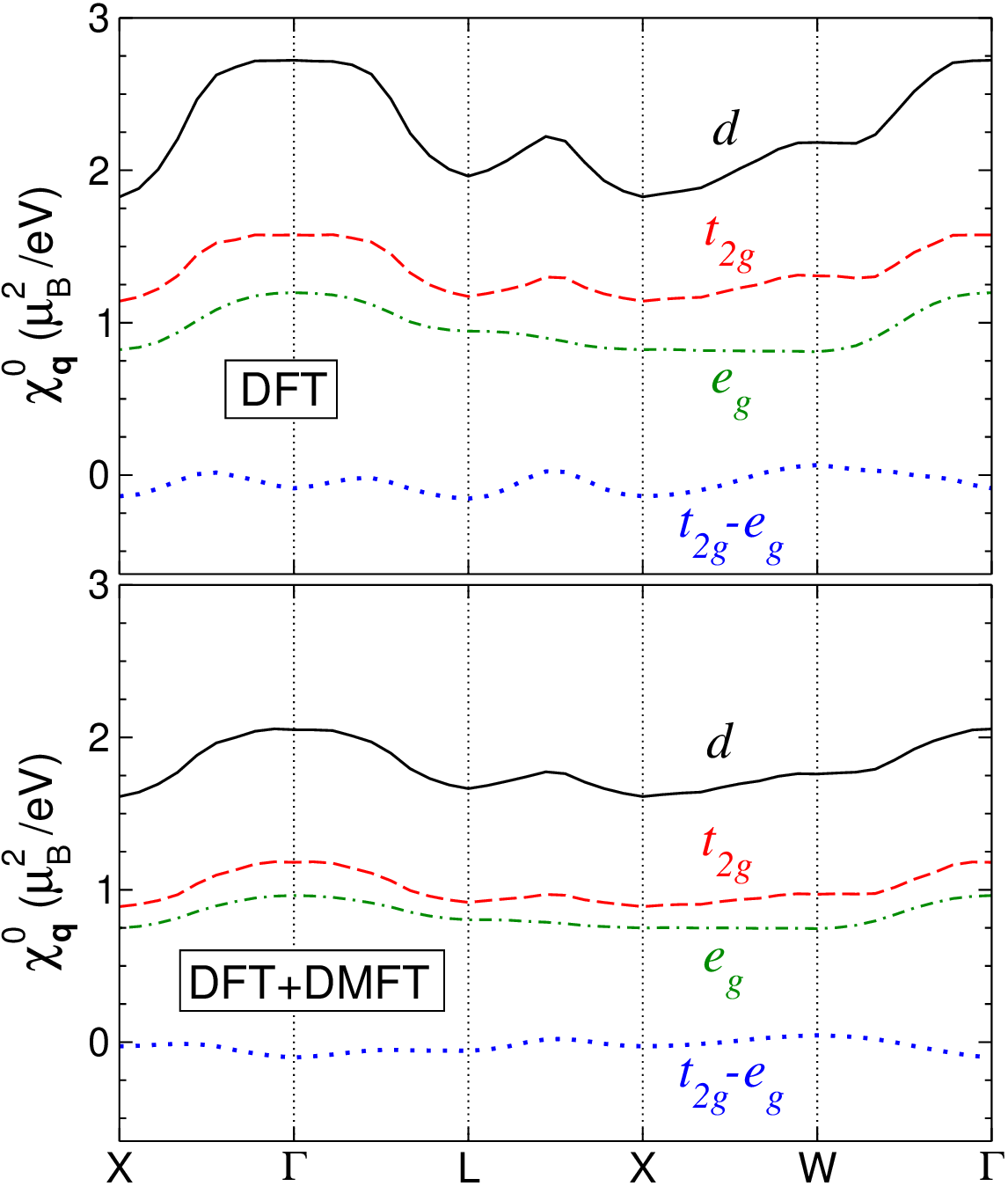}
\caption{
\label{fig:susc_irr}
Fig. 5. Momentum dependence of the static magnetic susceptibility for Co $3d$ states and its orbital-resolved contributions  obtained within DFT (top panel) and DFT+DMFT (bottom panel) at temperature ${T=1658}$~K.}
\end{figure}

To identify the dominant wave-vector of magnetic response,
we calculate the momentum dependence of the static magnetic susceptibility. This is done by considering the lowest-order contribution corresponding to the particle-hole bubble diagram:
\begin{equation}
\chi_{\bf q}^{0} = -\frac{2\mu_{\rm B}^2}{\beta} \sum_{{\bf k},\nu_n} \textrm{Tr}[G_{\bf k} (i\nu_n) G_{\bf k+q}(i\nu_n)].
\end{equation}
Here, $G_{\bf k}(i\nu_n)$ is the one-particle Green's function
at momentum ${\bf k}$, 
$\nu_n$ are Matsubara frequencies,
$\mu_{\rm B}^{}$ is the Bohr magneton and 
$\beta$ stands for the inverse temperature.

In Fig.~\ref{fig:susc_irr}
we present $\chi_{\bf q}^{0}$ for $3d$ states obtained using non-interacting (DFT) and interacting (DFT+DMFT) Green's functions.
One can see that the inclusion of dynamical electron correlations within DMFT does not qualitatively affect the behaviour of $\chi_{\bf q}^{0}$, but rather reduce its dependence on momentum. 
In both cases, $\chi_{\bf q}^{0}$ reaches a global maximum at the high-symmetry point $\Gamma$, corresponding to ferromagnetic ordering. 
Moreover, as seen from the orbital-resolved contributions to $\chi_{\bf q}^{0}$, both $t_{2g}$ and $e_g$ states have a maximum at the $\Gamma$~point.

{\bf Conclusion.}
We have investigated the electronic and magnetic properties of fcc cobalt by DFT+DMFT approach. 
The computed uniform and local magnetic susceptibilities follow the Curie-Weiss law, which has been shown to occur due to the partial formation of local magnetic moments. 
We found that the lifetime of these moments in cobalt is significantly less than in bcc iron, implying that the magnetism of cobalt is more itinerant.
Furthermore, contrary to previous reports for bcc iron~\cite{Katanin2010}, we have not observed substantial orbital selectivity in cobalt.
In particular, all obtained electronic self-energies in cobalt exhibit a quasiparticle shape with the quasiparticle mass enhancement factor ${m^*/m\sim 1.8}$, corresponding to moderately correlated metal.
Analyzing the momentum dependence of static magnetic susceptibility, we found a strong tendency to ferromagnetic ordering.

We thank A.~A.~Katanin for the valuable discussions.
The DFT+DMFT calculations were supported by the
Russian Science Foundation (project 19-12-00012).
A.S.B. was supported by the Alexander von Humboldt Foundation.


\begin{thebibliography}{99}
\bibitem{footnote1}
The DFT+DMFT calculations of local spin-spin correlation functions for chromium and bcc iron were performed at the same temperature ${T=1658}$~K as for cobalt. Other computational details can be found in Refs.~\cite{ourChromium,OurAlphaBelozerovKatanin}.
\end{thebibliography}
\end{document}